\documentclass[12pt,a4paper]{article}
\usepackage[latin1]{inputenc}
\usepackage{epsfig}
\usepackage{rotating}
\usepackage{fancyheadings}
\usepackage{isolatin1}
\title{Density waves in dry granular media falling through a vertical pipe}
\author{T. Raafat\(^{\star +}\), J.P. Hulin\(^{+}\), H.J. Herrmann\(^{\star}\)\\\\
\(^{\star}\)Laboratoire P.M.M.H., E.S.P.C.I., (URA CNRS \(\rm
n^{\circ}\) 857)\\
10 rue Vauquelin, 75231 Paris, Cedex 05 (France)\\
\(^{+}\)Laboratoire FAST, (URA CNRS \(\rm n^{\circ}\) 871)\\
Bat. 502, Campus Universitaire, 91405 Orsay Cedex (France)\\}
\date{}
\begin{document}
\maketitle
\begin{abstract}
We report experimental measurements of density waves
in granular materials flowing down in a capillary tube. The density
wave regime occurs at intermediate flow rates between a low density
free fall regime and a high compactness slower flow. We observe this
intermediate state when the ratio of the tube diameter and the
particles size lies between 6 and 30. The propagation velocity of the waves
is constant along the tube length and increases linearly with the
total mass flow rate \(\rm \Phi\). The wave structures include
compact clogs (lengths are independent of \(\rm \Phi\)) and bubbles of
low compactness (lengths increase with \(\rm \Phi\)). Both length
distributions are invariant along the tube length. A model assuming a
free fall regime in the bubbles and a compactness of \(35\%\) inside
the clogs allows to account for the mass distribution in the flow.
\end{abstract}
\newpage
\section{Introduction}
Nowadays dry granular media take up an important place in our life.
Scientists have studied these materials to understand its behavior in 
nature \cite{1}-\cite{17} and for industrial applications
\cite{3}\cite{4}\cite{17}. One of these problems, which we
are interested in understanding, is the appearance of density
waves in downward flows of granular media inside a pipe
\cite{17}. These effects have significant analogies with the traffic 
flow model that successfully characterizes traffic jams on highways 
\cite{1}. Several authors have already analyzed the problem 
\cite{1}-\cite{17}. However, the dependence of the
structure of the waves on the physical parameters controlling the flow
have not been studied systematically in these works. In the present 
letter we study in particular the evolution of the characteristic geometry
of the waves and of its propagation velocity in relation to the
total mass flow rate.
\section{Experimental setup}
The experimental setup (Fig.1) comprises a conical 
hopper with an opening angle of 60 degrees attached to a vertical
glass pipe of a length of 1.3 m and an internal
diameter of approximately 2.9 mm. At the bottom of the
pipe a variable closure of the outlet makes it possible to adjust the 
outflow. With an optical acquisition device we analyze variations of 
the grain packing fraction. Light of a standard 50 Watt halogen lamp 
falls onto a double slit which divides the light in two beams and then a
lens concentrates the light beams onto the pipe as two narrow
horizontal lines. Another lens
refocuses the light scattered and diffracted by the falling grains and
the pipe itself onto two light detection diodes on the other side. 
The high sensitivity of the diodes makes it
indispensable to protect the optical axis from stray radiation. 
Using two light beams allows to determine the average velocity of the 
density waves. We define two measurement heights, \(\rm h_{1}\) in the
vicinity of the top (20 cm below the hopper) and \(\rm h_{2}\) near
the bottom of the pipe (30 cm above the outlet), to study the
dependence of the variations of the grain density on the length of the
pipe. 
Simultaneously with the optical acquisition
we measure the mass flow as a function of the time by adding
electronic computer controlled scales under the outlet of the pipe. 
The time variations of the transmitted 
light corresponding to the various measured parameters are recorded and 
processed afterwards on a Unix Workstation.\\
We performed our experiments with small glass beads of a average
diameter of 200 \(\rm \mu m\) and small glass splinters of a mean size
between 90 and 200 \(\rm \mu m\).
\section{Qualitative observations}
Before performing the experiments,
it is important to avoid excess humidity in the granular materials,
else strong adhesive forces arise between the grains and
between grains and pipe. 
Simply blowing into the pipe creates enough humidity that the pipe can
only be used again after drying it with hot air.\\
The observed phenomena can be subdivided in three regimes:\\
The first and simplest case describes the behavior of grains falling 
down a pipe without or with only a small outlet closure: this case
corresponds to the largest mass flow rates.
When grains fall down from the hopper they drag air with them inducing
suction. In the mentioned case of largest mass flow rates 
this suction causes air to flow through the sand in the upper part of
the pipe. To verify this a balloon filled with air was put over the
hopper. Due to the suction the balloon contracted.\\
The second regime is characterized by a high compactness of the flowing
grains which is approximately constant over the length of the pipe. This
case corresponds to the lowest flow rate values (small outlet
opening).\\
Between the two cases mentioned above we observe a third regime
characterized by density waves. Each individual density wave
consists of two different parts. The first highly compact and dense
section will be called  a ``clog''. The second is a bubble filled with
air in which the particle density is lower and their velocity higher. 
When the outflow is reduced from the free fall case a plug builds up
at the bottom of the
pipe. In this case the air stream from the top to the bottom is
hindered by the plug. This assumption could also be verified with 
an air-balloon. The contraction of the balloon caused by the suction
in the free fall regime discussed above stops immediately with the
formation of the plug at the bottom of the pipe.\\  
Fig.2 illustrates a typical density wave of
particles. The radial structure of the bubbles has
been determined by video analysis.\\
To obtain the density waves it is necessary to keep the ratio rat of the 
diameter of the fixed pipe and the size of the particles within certain
limits. We have verified in several experiments the condition:
\begin{equation}
\rm 30\geq rat\geq 6.
\end{equation}
For \(\rm rat > 30\) we always obtain the free fall regime. 
For \(\rm rat < 6\) only the compact regime is observed and the flow
very often stops completely due to arching.\\
We have noticed that density waves do not appear directly at the
hopper outlet but at a certain distance \(\rm \Delta\) below. 
The distance \(\rm \Delta\) increases with grain size.
\section{Analysis of density variations using light transmission}
As mentioned above we analyze the time variations of the intensity of 
light transmitted through the pipe. Considering typical variation 
in the regime of density waves the high intensity peaks correspond to 
air bubbles and the intervals between the peaks represent clogs. To extract
information from these time series we have introduced a threshold
distinguishing between high (clogs) and low (air bubbles)
compactness. We have chosen this threshold value just large enough to
eliminate the influence of the noise of the base line as one can see
in Fig.3.\\
We obtain a binary curve by replacing the data points above the
threshold by one and the others by zero. In this way we introduce for
each individual bubble and clog labeled i the respective
characteristic transit times \(\rm \tau_{x}^{i}\) and 
\(\rm\tau_{l}^{i}\). They are defined as the time during which the
measured signal remains respectively higher or lower than the
threshold when the bubble or the clog moves through the measurement
section. Plotting the different characteristic transit times as a
function of time leads to histograms. The histogram of the duration of
the clogs displays a well defined peak
corresponding to the average transit time \(\rm \tau_{l}\) of the
clogs. In contrast the transit time distribution of the bubbles is
much broader, so that we obtain \(\rm \tau_{x}\) by averaging over all
\(\rm \tau_{x}^{i}\). Fig.4 shows some typical histograms. 
We have superimposed two measurements performed at two different heights 
with approximately the same mass flow. We observe the changes with
distance of the histograms are extremely small: this implies that the
granular flow has reached a stationary regime even at the upper
measurement level.\\
Measurements using two light beams displaced a distance D on the axis of
the pipe (Fig.1) lead to two similar time series, shifted a time \(\rm
\tau_{D}\) representing the transit time of the density variations
between the two measurement sections. 
From the peak of the correlation function of the two time series we
obtain \(\rm\tau_{D}\) and thus the apparent velocity \(\rm v_{l} =
D/\tau_{D}\) of the clogs.
In table 1 and 2, concerning the
different measurement heights \(\rm h_{1}\) and \(\rm h_{2}\), one can
see the relations between \(\rm\tau_{l},\,\tau_{x},\,\tau_{D}\) and
the total mass flow \(\rm \Phi\). Furthermore we see also the relation
between the transit times, the velocity \(\rm v_{l}\) and the calculated 
values for the respective lengths x of the bubbles and l of the clogs. 
The plot of the global
mass flow rate \(\rm \Phi\) as a function of \(\rm v_{l}\) in Fig.5 shows an
almost linear increase of \(\rm \Phi\) with \(\rm v_{l}\), so that we
can write:
\begin{equation}
\rm \Phi = \Phi_{0} + A\,v_{l}.
\end{equation}
We obtain that \(\rm A=4.9\, g/m\) and \(\rm\Phi_{0}=1.2\,
g/s\).\\
We estimate the characteristic length l of the clogs from their
velocity and transit time,
\begin{equation}
\rm l = v_{l} \, \tau_{l}.
\end{equation}
In Fig.5 we have superimposed two series of measurements performed
at two different measurement heights \(\rm h=h_{1}\) and \(\rm
h=h_{2}\). We observe that
the velocity \(\rm v_{l}\) is independent of the measurement heights 
h and therefore we assume that \(\rm v_{l}\) is constant
along the pipe. We calculate the average length x of the low particle 
density bubbles from the mean transit time  \(\rm\tau_{x}\) of a
bubble through a light beam.
\[\rm v \simeq v_{l} = const,\]
\begin{equation}
\rm \frac{l}{\tau_{l}} = \frac{x}{\tau_{x}}.
\end{equation}
Eq.(4) leads to
\begin{equation}
\rm x = l \, \frac{\tau_{x}}{\tau_{l}}.
\end{equation}
Fig.6 illustrates the relation between l, x and the
global mass flow rate \(\rm \Phi\). The two different kinds of symbols
represent the two different measurement heights mentioned above.\\
As can be seen in Fig.6 the length l of the clogs is nearly
independent of 
both the mass flow rate and the height at which the
density measurement is performed. In contrast, the average length x of
the air bubbles varies. We observe at both the higher and lower
measurement height an almost linear increase of x
with \(\rm \Phi\).\\
After computing the respective lengths l and x it is possible to
obtain the mean masses \(\rm m_{l}\) and \(\rm m_{x}\) of the granular
material in a clog and in a bubble. In order to estimate \(\rm m_{l}\)
we have measured
independently the total mass \(\rm m_{p}\) of the grains filling
completely the full length \(\rm L_{p}\) of the pipe under zero flow
conditions. \(\rm m_{l}\) verifies:
\begin{equation}
\rm m_{l} = \frac{m_{p}}{L_{p}} \, l \, \frac{c(\Phi)}{c_{0}}.
\end{equation}
The variable \(\rm c(\Phi)\) represents the compactness of the grain 
packing in the flowing clogs and \(\rm c_{0}\) in a non flowing 
packing corresponding to the mass \(\rm m_{p}\) (experimentally, one
finds: \(\rm c_{0} \approx 0.63\)).\\
In the next step we calculate the mean mass \(\rm m_{x}\) of a low
density pocket. Let us consider now the mass flow rate of grains \(\rm
\Phi_{v,g}\) in an inertial frame moving at the velocity \(\rm
v_{l}\) of the clog. The density wave structure in this reference frame 
is stationary. This assumption results in a total mass flow rate \(\rm
\Phi\) in the fixed reference laboratory frame of
\begin{equation}
\rm \Phi = \Phi_{v,g} + \Phi_{v,l},
\end{equation}
with
\begin{equation}
\rm \Phi_{v,l} = \frac{m_{x} + m_{l}}{\tau_{x} + \tau_{l}} = v_{l} \, \frac{m_{x} + m_{l}}{x + l}.
\end{equation}
The initial velocity \(\rm v_{0,g}\) of the grains at the
top of an air bubble in the moving reference frame is assumed to
verify
\begin{equation}
\rm v_{0,g} = \frac{\Phi_{v,g}}{\pi \, r^{2} \,
  \rho_{g} \, c(\Phi)} = \frac{\Phi_{v,g}}{K \, c(\Phi)},
\end{equation}
with
\begin{equation}
\rm K = \pi \, r^{2} \, \rho_{g}, 
\end{equation}
\(\rm \rho_{g}\) is the bulk density of the glass used for the
grains. Equation (9) implies that both the grain concentration and
their velocity are continuous at the bottom of the clogs. The unknown 
variables are \(\rm m_{x}\), \(\rm \Phi_{v,g}\) and \(\rm c(\Phi)\).
We assume that the grains in an air bubble fall freely and the 
interactions between the grains and the particles of the wall are
negligible, so that
\begin{equation}
\rm \frac{d v_{g}}{d t} = g.
\end{equation}
The grain velocity \(\rm v_{g}\) in the reference moving frame 
is related to \(\rm \Phi_{v,g}\),
\begin{equation}
\rm \Phi_{v,g} = \rho(z)\,v_{g}(z),
\end{equation}
in which \(\rm \rho(z)\) is the integration of the local mass density 
over the cross section at height z.
One obtains the mass of particles \(\rm m_{x}\) inside a bubble
using relations (11) and integrating \(\rm \rho(z)\) from 0 to x to
compute the grain mass density:
\begin{equation}
\rm m_{x} = \int_{0}^{x} \rho(z)\,dz =
\frac{\Phi_{v,g}^{2}}{g\,K\,c(\Phi)}\,(\sqrt{1 +
  \frac{2\,x\,g\,K^{2}\,c^{2}(\Phi)}{\Phi_{v,g}^{2}}} - 1). 
\end{equation}
Inserting Eq.(13) into Eq.(7) leads to
\begin{equation}
\rm \Phi = \Phi_{v,g} + \frac{v_{l}}{x + l}\left(\frac{m_{p}}{L_{p}} \, l \, \frac{c(\Phi)}{c_{0}} +
\frac{\Phi_{v,g}^{2}}{g\,K\,c(\Phi)}\left(\sqrt{1 + \frac{2\,x\,g\,K^{2}\,c^{2}(\Phi)}{\Phi_{v,g}^{2}}} - 1\right)\right). 
\end{equation}
We solve Eq.(14) numerically to obtain \(\rm\Phi_{v,g}\) under
the assumption \(\rm c(\Phi) = const\). We used the criterion that
when \(\rm v_{l} = 0\) the total flux equals the grain flow rate, i.e.,
\begin{equation}
\rm \Phi(v_{l}=0) = \Phi_{v,g}(v_{l}=0),
\end{equation}
to obtain \(\rm c(\Phi)\)= 0.35. We see in table 3 and 4,
corresponding to the different measurement
heights \(\rm h_{1}\) and \(\rm h_{2}\), that \(\rm\Phi_{v,g}\) and
\(\rm\Phi_{v,l}\) increase with the velocity \(\rm v_{l}\), but
\(\rm\Phi_{v,l}\) remains for each value of the velocity under \(50\%\)
of the value of \(\rm\Phi_{v,g}\). We have illustrated these relations
in Fig.5. The upper line corresponds to the total mass flow rate and
the others correspond to \(\rm\Phi_{v,g}\) and \(\rm\Phi_{v,l}\).\\
Replacing \(\rm\Phi_{v,g}\) in Eq.(13) gives us the masses 
\(\rm m_{x}\) for the measurement heights \(\rm h_{1}\) and \(\rm
h_{2}\). Decreasing the total mass flow rate \(\rm \Phi\) leads to an
approach of the masses \(\rm m_{x}\) and \(\rm m_{l}\). This behavior
is consistent with the lengths of the bubbles and the clogs as one can
see in Fig.6.
\section{Conclusion}
In the present letter we have verified the existence of a regime of density
waves in dry granular media at intermediate flow rates between a low
density free fall regime and a slow regime of flow of high
compactness. We obtain nearly stationary structures of waves including
compact clogs and bubbles of low compactness. Both the length distributions
of clogs and bubbles and the propagation velocity of the waves are
constant along the pipe. The lengths of the clogs are independent of
the total mass flow rate \(\rm \Phi\). However, the lengths of the
bubbles increase with \(\rm \Phi\) and they are a factor 1.5 to 3
larger than the clog lengths. A model assuming a free fall inside the 
bubbles and a compactness of \(35\%\) inside the clogs well reproduces
the mass distribution in the flow.\\ It is clear that there are
still numerous problems which have to be solved. A key issue for the
full understanding of the phenomena will be the determination of the
forces acting on the grains both in the clog and bubble zones. This
includes considering friction forces between the grains and the walls
and further hydrodynamic forces resulting from the gas in the column.
A crucial point will be the determination and prediction of the
pressure gradients in the column.
\newpage
\begin{center}
{\Large\bf Acknowledgment}\\
\end{center}
The authors thank gratefully S. Schwarzer and H. Puhl for a critical 
reading of the manuscript.

independence of the measurement height. Assuming that the compactness
\newpage{
\vspace*{1cm}
\begin{center}
{\Large\bf Figures}
\end{center}
\vspace{1cm}
\begin{center}
\begin{itemize}
\item Figure 1: Experimental Setup.\\
\vspace{1cm}
\item Figure 2: Schematic view of density waves formed by particles.\\
\vspace{1cm}
\item Figure 3: Typical time variation of the intensity of light
transmitted  through the pipe with defined threshold to
distinguish between high and low compactness (\(\rm \Phi=1.980\) g/s).\\
\vspace{1cm}
\item Figure 4: Histograms of time for clogs and air bubbles in a semi-log
plot. The continuous lines correspond to the measurement height
\(\rm h_{2}\), with \(\rm \Phi(h_{2})=2.466\) g/s. The dotted
lines correspond to the measurement height \(\rm h_{1}\), with
\(\rm \Phi(h_{1})=2.459\) g/s.\\
\vspace{1cm}
\item Figure 5: Total mass flow as a function of measured clog velocity
along the pipe. The triangles correspond to the measurement height
\(\rm h_{1}\) and the checks to the measurement height \(\rm h_{2}\).\\
\vspace{1cm}
\item Figure 6: Characteristic lengths of clogs (lower values) and
air bubbles (upper values) as a function of the mass flow. The triangles
correspond to the measurement height \(\rm h_{1}\) and the checks
to the measurement height \(\rm h_{2}\).\\
\end{itemize}
\end{center}}
\newpage{
\vspace*{1cm}
\begin{center}
{\Large\bf Tables}
\end{center}
\vspace{1cm}
\begin{center}
\begin{itemize}
\item Table 1: \(\rm \tau_{l},\,\tau_{x},\,\tau_{D},\,v_{l},\,x,\,l\) as a
function of the total mass flow \(\rm \Phi\) (measurement height
\(\rm h_{1}\)).\\
\vspace{1cm}
\item Table 2: \(\rm \tau_{l},\,\tau_{x},\,\tau_{D},\,v_{l},\,x,\,l\) as a
function of the total mass flow \(\rm \Phi\) (measurement height
\(\rm h_{2}\)).\\
\vspace{1cm}
\item Table 3: \(\rm
\Phi,\,\Phi_{v,g},\,\Phi_{v,l},\,v_{l},\,m_{l},\,m_{x}\)
(measurement height \(\rm h_{1}\)).\\
\vspace{1cm}
\item Table 4: \(\rm
\Phi,\,\Phi_{v,g},\,\Phi_{v,l},\,v_{l},\,m_{l},\,m_{x}\)
(measurement height \(\rm h_{2}\)).\\
\end{itemize}
\end{center}}

\newpage

\begin{figure}[htbp]
  \begin{center}
    \leavevmode
    \hspace{-1cm}\epsfig{file=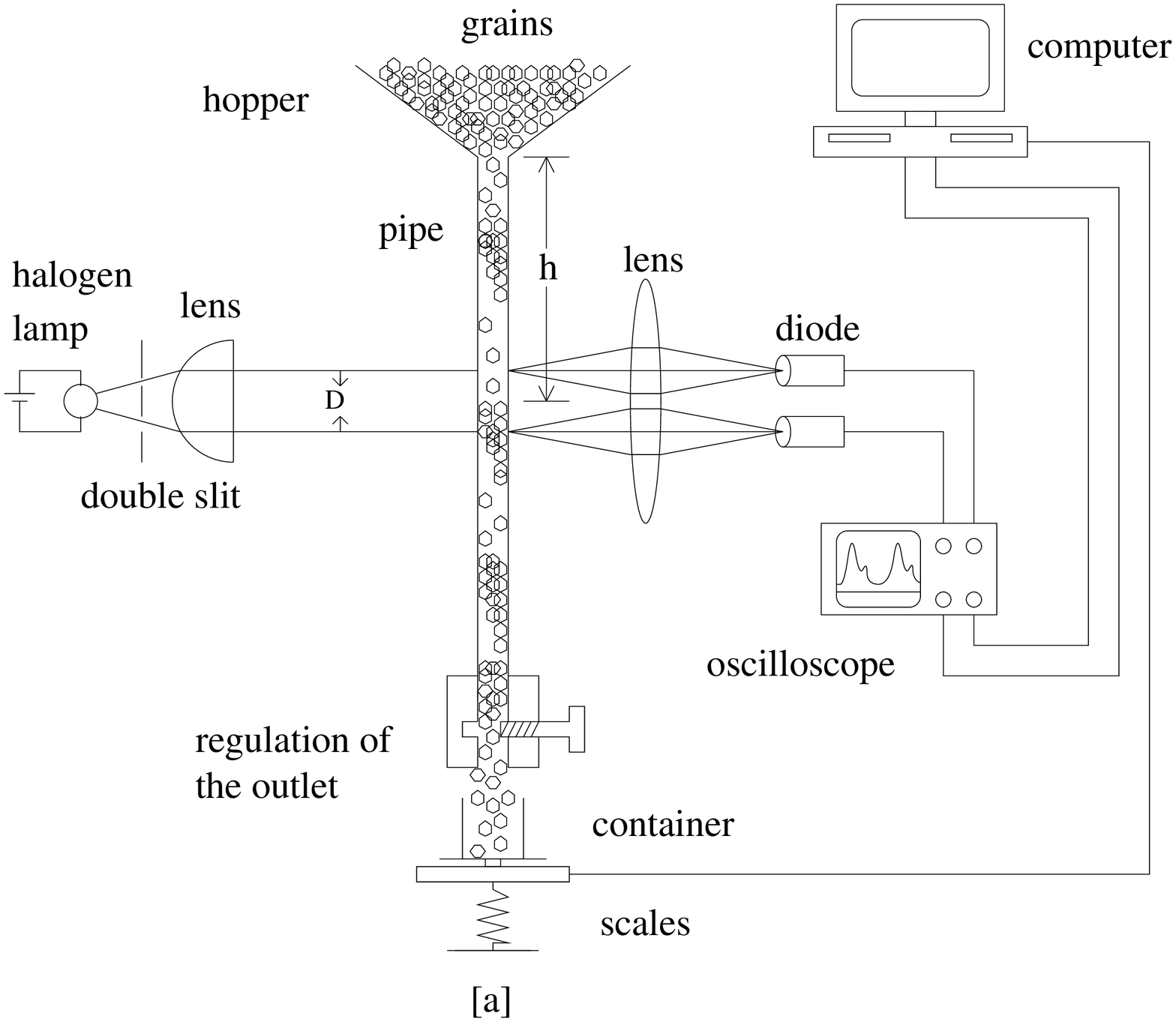,width=\textwidth}
  \end{center}
  \caption{}
\end{figure}

\newpage

\begin{figure}[htbp]
  \begin{center}
    \leavevmode
    \hspace{-1cm}\epsfig{file=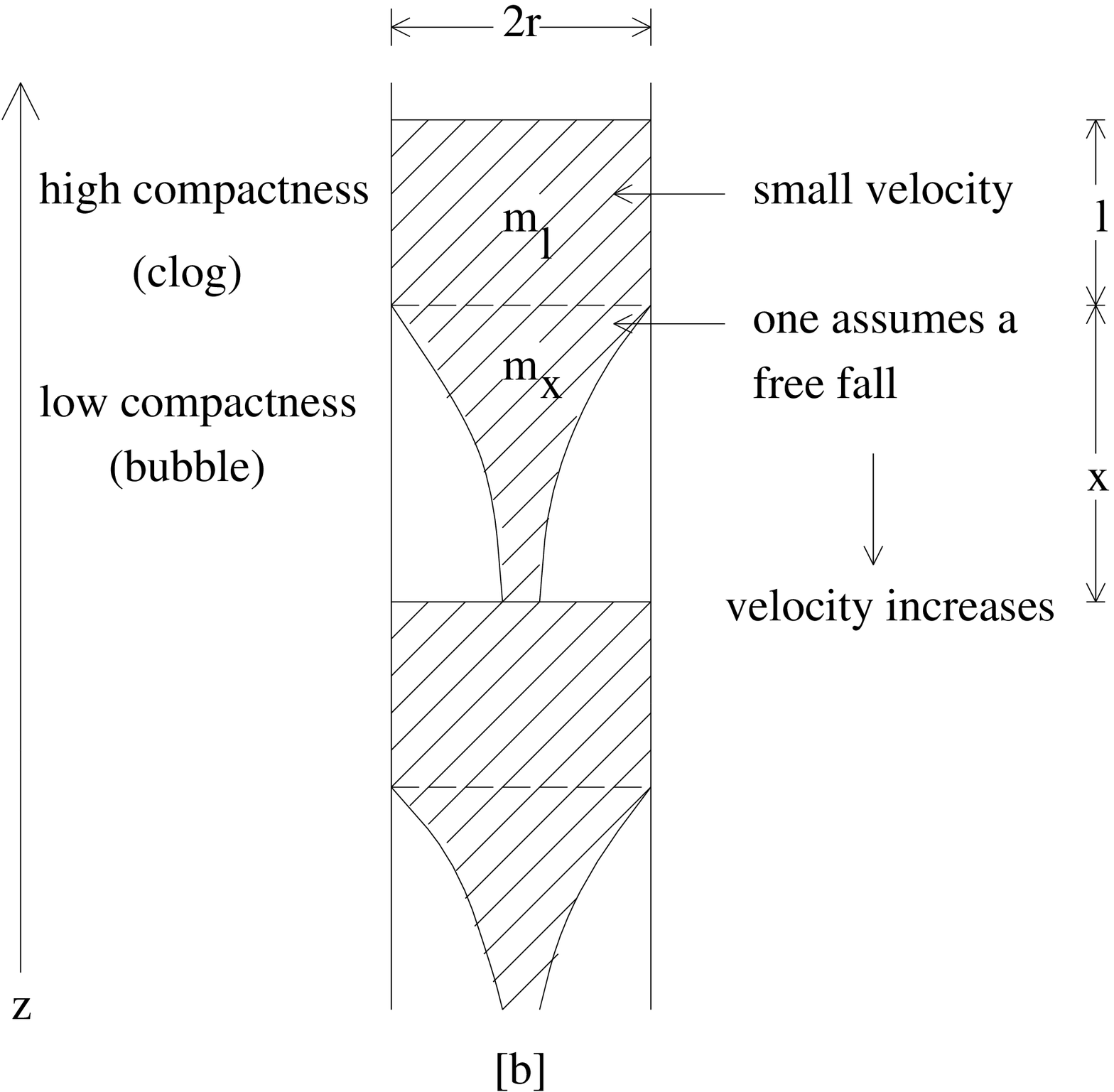,width=\textwidth}
  \end{center}
  \caption{}
\end{figure}

\newpage

\begin{figure}[htbp]
  \begin{center}
    \leavevmode
    \hspace{-1cm}\epsfig{file=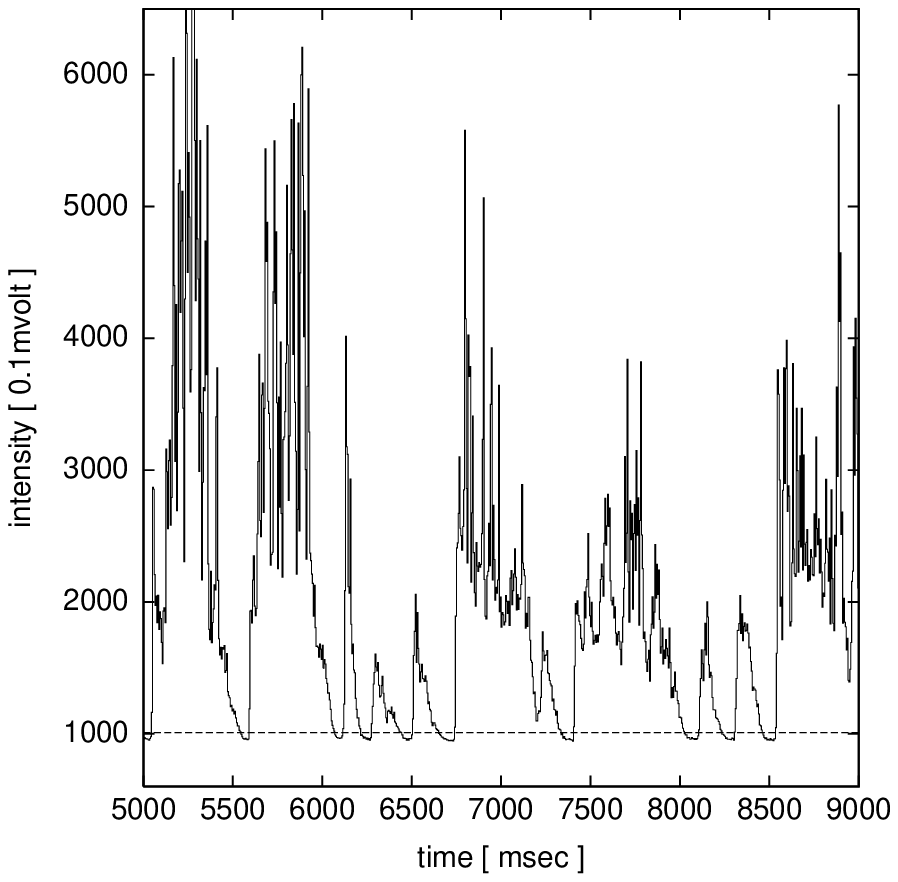,width=\textwidth}
  \end{center}
  \caption{}
\end{figure}

\newpage

\begin{figure}[htbp]
  \begin{center}
    \leavevmode
    \hspace{-1cm}\epsfig{file=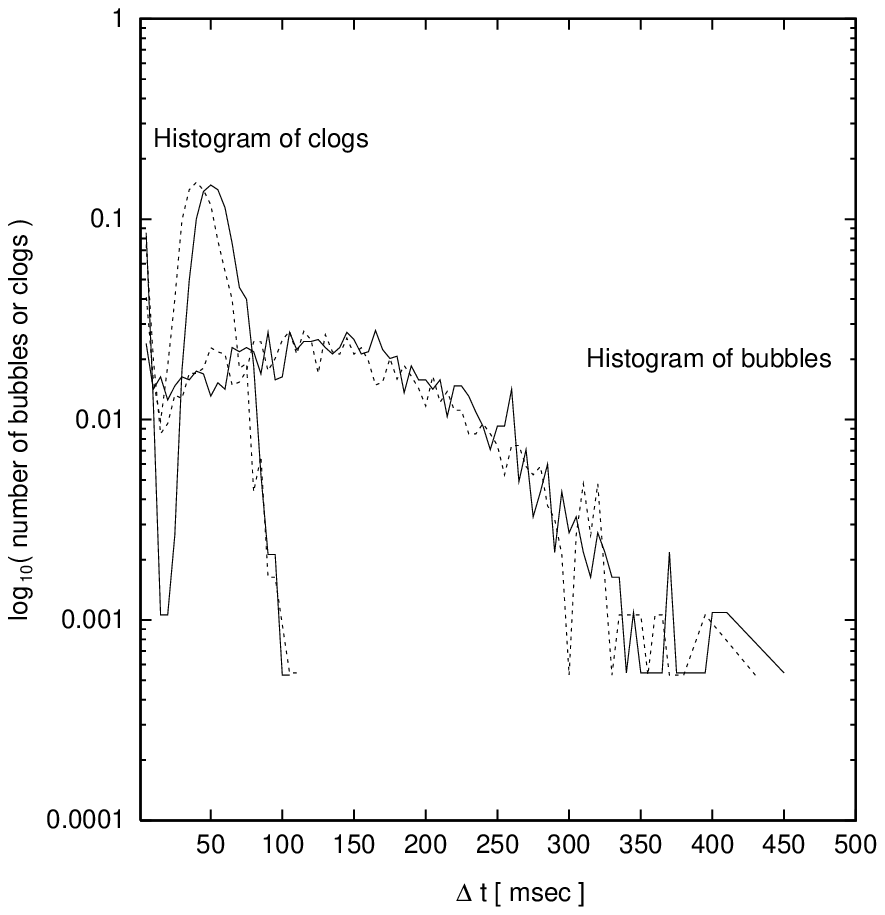,width=\textwidth}
  \end{center}
  \caption{}
\end{figure}

\newpage

\begin{figure}[htbp]
  \begin{center}
    \leavevmode
    \hspace{-1cm}\epsfig{file=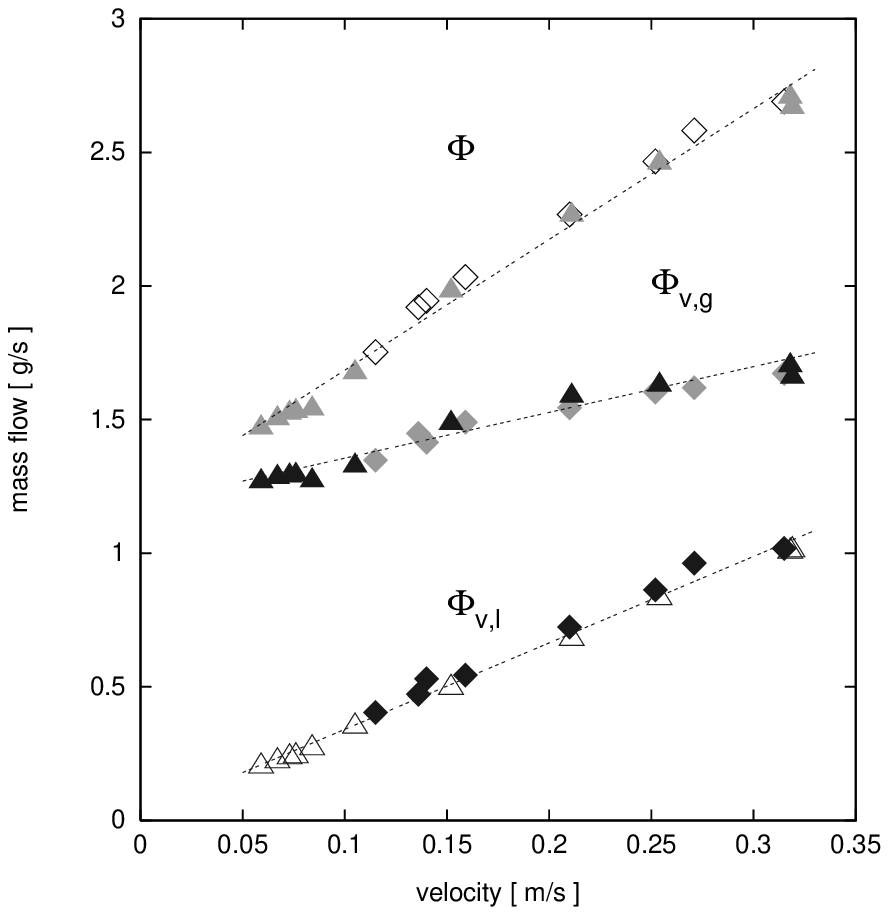,width=\textwidth}
  \end{center}
  \caption{}
\end{figure}

\newpage

\begin{figure}[htbp]
  \begin{center}
    \leavevmode
    \hspace{-1cm}\epsfig{file=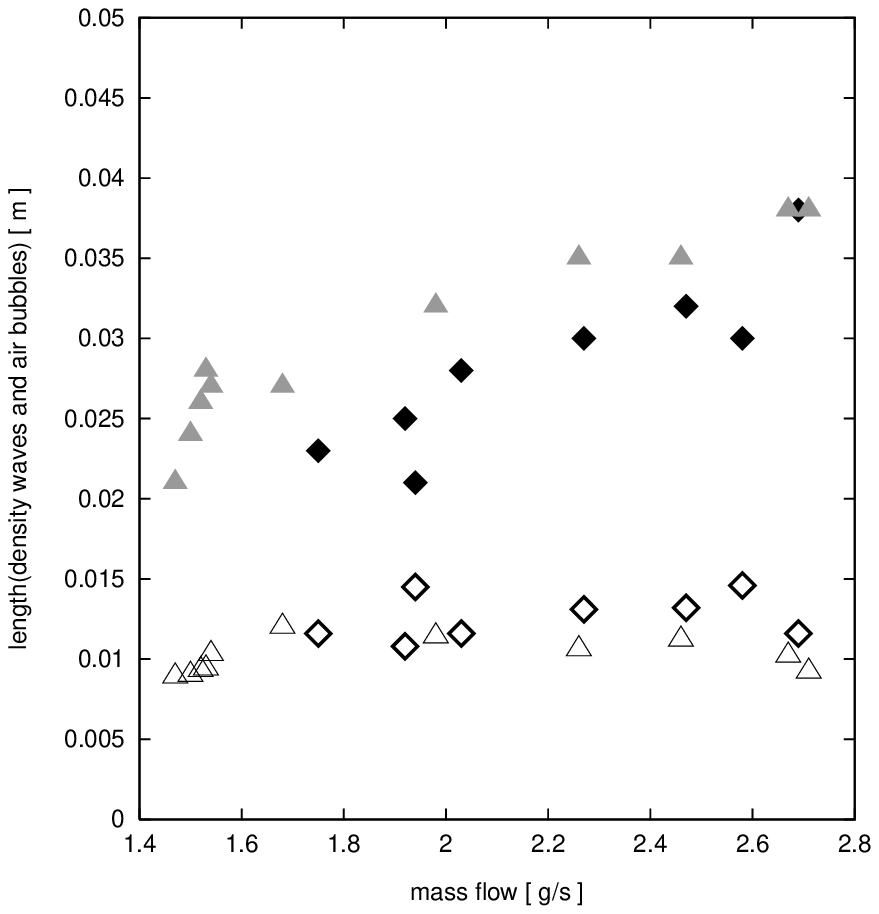,width=\textwidth}
  \end{center}
  \caption{}
\end{figure}

\newpage{

\begin{table}\centering
\begin{tabular}{|c|c|c|c|c|c|c|c|c|}
\hline
{$\rm\Phi\,[g/s]$}&2.67&2.71&2.46&2.26&1.98&1.68&1.54&1.53\\
\hline
{$\rm\tau_{l}\,[ms]$}&32.09&28.89&44.22&50.42&75.24&114.24&122.22&123.75\\
\hline
{$\rm\tau_{x}\,[ms]$}&118.81&118.75&136.38&167.75&212.96&260.77&317.51&369.28\\
\hline
{$\rm\tau_{D}\,[ms]$}&59.56&59.75&74.80&90.05&13&180.95&226.19&250\\
\hline
{$\rm v_{l}\,[m/s]$}&0.319&0.318&0.254&0.211&0.152&0.105&0.084&0.076\\
\hline
{$\rm x\,[m]$}&0.038&0.038&0.035&0.035&0.032&0.027&0.027&0.028\\
\hline
{$\rm l\,[m]$}&0.010&0.009&0.011&0.011&0.011&0.012&0.010&0.009\\
\hline
\end{tabular}
  \caption{}
\end{table}}

\newpage{

\begin{table}\centering
\begin{tabular}{|c|c|c|c|c|c|c|c|c|}
\hline
{$\rm\Phi\,[g/s]$}&2.69&2.58&2.47&2.27&2.03&1.94&1.92&1.75\\
\hline
{$\rm\tau_{l}\,[ms]$}&36.70&53.86&52.56&62.48&72.86&103.35&79.04&101.05\\
\hline
{$\rm\tau_{x}\,[ms]$}&120.07&110.77&125.31&144.15&177.91&152.75&188.43&199.71\\
\hline
{$\rm\tau_{D}\,[ms]$}&60.32&70.11&75.40&90.48&119.50&135.71&139.71&165.22\\
\hline
{$\rm v_{l}\,[m/s]$}&0.315&0.271&0.252&0.210&0.159&0.140&0.136&0.115\\
\hline
{$\rm x\,[m]$}&0.038&0.030&0.032&0.030&0.028&0.021&0.025&0.023\\
\hline
{$\rm l\,[m]$}&0.012&0.015&0.013&0.013&0.012&0.015&0.011&0.012\\
\hline
\end{tabular}
  \caption{}
\end{table}}

\newpage{

\begin{table}\centering
\begin{tabular}{|c|c|c|c|c|c|c|c|c|}
\hline
{$\rm\Phi\,[g/s]$}&2.67&2.71&2.46&2.26&1.98&1.68&1.54&1.53\\
\hline
{$\rm\Phi_{v,g}\,[g/s]$}&1.66&1.70&1.63&1.59&1.49&1.33&1.27&1.29\\
\hline
{$\rm\Phi_{v,l}\,[g/s]$}&1.01&1.01&0.83&0.68&0.50&0.35&0.27&0.24\\
\hline
{$\rm v_{l}\,[m/s]$}&0.319&0.318&0.254&0.211&0.152&0.105&0.084&0.076\\
\hline
{$\rm m_{l}\,[g]$}&0.052&0.047&0.057&0.054&0.058&0.061&0.052&0.048\\
\hline
{$\rm m_{x}\,[g]$}&0.101&0.103&0.094&0.093&0.084&0.069&0.067&0.070\\
\hline
\end{tabular}
  \caption{}
\end{table}}

\newpage{

\begin{table}\centering
\begin{tabular}{|c|c|c|c|c|c|c|c|c|}
\hline
{$\rm\Phi\,[g/s]$}&2.69&2.58&2.47&2.27&2.03&1.94&1.92&1.75\\
\hline
{$\rm\Phi_{v,g}\,[g/s]$}&1.67&1.62&1.60&1.54&1.49&1.42&1.45&1.35\\
\hline
{$\rm\Phi_{v,l}\,[g/s]$}&1.02&0.96&0.86&0.72&0.54&0.53&0.47&0.40\\
\hline
{$\rm v_{l}\,[m/s]$}&0.315&0.271&0.252&0.210&0.159&0.140&0.136&0.115\\
\hline
{$\rm m_{l}\,[g]$}&0.059&0.074&0.067&0.066&0.059&0.074&0.055&0.059\\
\hline
{$\rm m_{x}\,[g]$}&0.102&0.085&0.088&0.082&0.077&0.061&0.070&0.063\\
\hline
\end{tabular}
  \caption{}
\end{table}}
\end{document}